\documentclass[twocolumn]{aastex631}

\newcommand{\teff}{$T_{\rm eff}$}

\newcommand{\logg}{$\log g$}

\newcommand{\feh}{$\rm{[Fe/H]}$}

\newcommand{\vmic}{$v_{t}$}

\newcommand{\qq}{$\mathrm{q}^{2}$}

\newcommand{\sm}{$\rm{M_{\odot}}$}

\newcommand{\be}{\begin{equation}}
\newcommand{\ee}{\end{equation}}
\newcommand{\ben}{\begin{eqnarray}}
\newcommand{\een}{\end{eqnarray}}
\newcommand{\bfg}{\begin{figure}}
\newcommand{\efg}{\end{figure}}

\newcommand{\li}{$A$(Li)}

\newcommand{\g}{$G$}

\accepted{\today}
\date{Accepted XXX. Received YYY; in original form ZZZ}

\shorttitle{Beryllium}
\shortauthors{Reggiani et al.}

\begin{document}

\title{Constraining Extra Mixing during the Main Sequence: Whatever Depletes
Lithium Does Not Touch Beryllium\footnote{Based on observations collected at the European Southern Observatory under ESO programme 111.24VX.}}

\correspondingauthor{H. Reggiani}
\email{henrique.reggiani@noirlab.edu}

\author[0000-0001-6533-6179]{Henrique Reggiani}
\affiliation{Gemini South, Gemini Observatory, NSF's NOIRLab, Casilla 603, La Serena, Chile}

\author[0000-0001-9261-8366]{Jhon Yana Galarza}
\altaffiliation{Carnegie Fellow}
\affiliation{The Observatories of the Carnegie Institution for Science, 813 Santa Barbara Street, Pasadena, CA 91101, USA}

\author[0000-0002-1387-2954]{Diego Lorenzo-Oliveira}
\affiliation{Laborat\'orio Nacional de Astrof\'isica, Rua Estados Unidos 154, 37504-364, Itajubá - MG, Brazil}

\author[0000-0003-1858-561X]{Sofia Covarrubias}
\affiliation{Department of Astronomy, California Institute of Technology, 1200 E. California Blvd., Pasadena, CA 91125, USA}
\affiliation{The Observatories of the Carnegie Institution for Science, 813 Santa Barbara Street, Pasadena, CA 91101, USA}

\author[0009-0006-8679-6282]{Micaela Oyague}
\affiliation{Seminario Permanente de Astronomía y Ciencias Espaciales, Facultad de Ciencias Físicas, Universidad Nacional Mayor de San Marcos, Avenida Venezuela s/n, Lima 15081, Perú}

\author{Rita Valle}
\affiliation{Seminario Permanente de Astronomía y Ciencias Espaciales, Facultad de Ciencias Físicas, Universidad Nacional Mayor de San Marcos, Avenida Venezuela s/n, Lima 15081, Perú}

\author[0000-0003-2481-4546]{Julio Chanamé}
\affiliation{Instituto de Astrof\'isica, Pontificia Universidad Cat\'olica de Chile, Av. Vicu\~na Mackenna 4860, 782-0436 Macul, Santiago, Chile}

\begin{abstract} 
Measurements of lithium abundances in solar-type stars have shown that standard models of stellar evolution are incapable of explaining the observed depletion as a function of stellar age. Beryllium is one of the lightest elements that can be measured in stellar photospheres, and it can be burned in relatively low temperatures. Studying its abundances as a function of stellar age can provide important constraints to stellar mixing models, as the level of depletion as a function of time will indicate how deep the photospheric material must be dredged to explain the observed abundances. In an effort to provide the most stringent constraints for non-standard stellar mixing models, we observed a sample of solar-twins and concomitantly analyzed their lithium and beryllium abundances. Unlike what is typically observed for lithium, we found that beryllium does not decrease as a function of stellar age along the main-sequence, constraining models that predict burning of both materials. Based on our data, models that invoke convective overshoot and convective settling are preferred over typical rotationaly-induced mixing models, as the later burn Be in excess while the former do not. Previous works also proposed mixing due to gravity waves as a possible explanation for observed abundances, which can fit our data as well. Furthermore, based on our solar twins, Be depletion likely happens within the first $\sim1$ Gyr. We also confirm previous findings of an increase in Be abundance as a function of metallicity, indicative of galactic production via cosmic ray spallation.
\end{abstract}

\keywords{Spectroscopy (1558) --- Stellar abundances (1577) --- Stellar atmospheres (1584) --- Fundamental parameters of stars (555) --- Solar analogs (1941)} 

\section{Introduction}\label{sec:intro}

The two lightest metals in the Universe are beryllium (Be) and lithium (Li). These fragile atoms are burned at relatively low temperatures ($2.5\times10^{6}$ and $3.5\times10^6$ K, Li and Be respectively). In the Sun, Li burning requires temperatures that are somewhat higher than those reached at the base of the convective zone, according to standard models of stellar evolution \citep[e.g.,][]{Schwarzschild:1957ApJ...125..233S, Weymann:1965ApJ...142..174W,Pinsonneault:1997ARA&A..35..557P}. Nevertheless, the observed photospheric abundances of Li (estimated using the most up-to-date methodology, i.e., three-dimensional and non-local thermodynamic equilibrium, non-LTE, modeling) are considerably lower than those inferred from the analysis of meteorites \citep[$0.96 \times 3.25$ for the Solar Photosphere vs. Meteorites, ][]{Greenstein:1951ApJ...113..536G,Weymann:1965ApJ...142..174W, Grevesse:2019BSRSL..88....5G, asplund2021}, which are representative of the Li composition of the Sun at the Zero Age Main Sequence (ZAMS). Therefore, there must have been lithium burning after the Sun entered the main-sequence. Which leads to the conclusion that there must be physical mechanisms responsible for transporting Li deeper than the base of the convective zone, where temperatures are high enough for it to be destroyed during the main-sequence.

Different mechanisms have been proposed to explain this ``extra mixing", such as rotation \citep[e.g.][]{Pinsonneault:1989ApJ...338..424P}; internal gravity waves \citep[e.g.,][]{Charbonnel:2005Sci...309.2189C}; microscopic diffusion and gravitational settling \citep[e.g.,][]{Michaud:2004ApJ...606..452M}; convective overshooting \citep[e.g.,][]{Xiong:2007ChA&A..31..244X, Zhang:2019ApJ...881..103Z}; and convective settling \citep[e.g.,][]{Andrassy:2015A&A...579A.122A}. The mechanism(s) responsible for the observed extra mixing could be one of or, more likely, a combination of, these proposed candidates. This extra mixing is clearly observed in the analysis of solar twins, in which there is a correlation between lithium abundance and main-sequence lifetime \citep[e.g.,][]{Carlos:2016A&A...587A.100C,Carlos:2019MNRAS.tmp..667C, boesgaard2022, Martos:2023MNRAS.522.3217M}. Lithium depletion has been observed in early main-sequence stars of other spectral types as well \citep[e.g.,][]{Boesgaard:1986ApJ...302L..49B}.  \citet{Thevenin:2017A&A...598A..64T} proposed that Li depletion in solar twins can be explained through pre-main sequence burning. However, \citet{Carlos:2019MNRAS.tmp..667C}, \citet{Martos:2023MNRAS.522.3217M} and \citet{Rathsam:2023MNRAS.525.4642R}, using large samples of solar twins and solar analogs, confirmed the existence of a correlation between Li abundances and stellar ages ($\tau$) for solar twins, a strong evidence of main-sequence depletion.

After lithium, beryllium is the easiest element to burn in stellar interiors. Its depletion happens at slightly higher temperatures, and measurements of its abundance could help distinguish between the different models put forward to explain the stellar lithium depletion, and the extent of this extra mixing that we observe. However, Be abundances are notoriously difficult to measure \citep[e.g.,][]{Balachandran:1998Natur.392..791B, takeda2011, asplund2021,korotin2022, boesgaard2022, Boesgaard:2023ApJ...943...40B}. The observable beryllium transitions lie in the UV region of the spectrum (3130, and 3131 \AA), a region poorly studied (when compared to optical wavelength regions), where uncertainties in atomic data are higher, and there is a wealth of lines and transitions not ideally characterized, with a large number of both molecular and atomic transitions contributing to the difficulties of Be measurements. In the region where the beryllium transitions are observed, there are also historically unaccounted for (until recently) UV opacities \citep{Balachandran:1998Natur.392..791B, Bell:2001ApJ...546L..65B, Short:2009ApJ...691.1634S}. Due to all the challenges surrounding the measurements of this element, it has historically been a point of contention. 

Early works suggested Be was depleted in the solar photosphere, compared to meteorites \citep{Chmielewski:1975A&A....42...37C}. The inconsistency between solar photospheric beryllium and the meteoritic value remained until recently. Using more up-to-date analysis methodologies, \citet{korotin2022} showed that the abundance of Be in the solar photosphere matches those inferred from meteorites \citep[A(Be)$=1.32\pm0.05$ dex for the solar photosphere and A(Be)$=1.31\pm0.04$ dex for meteorites,][]{korotin2022,Lodders:2021SSRv..217...44L}.

Using beryllium as a secondary probe of stellar mixing is not new in the literature, an ongoing effort since the discovery by \cite{Boesgaard:1986ApJ...302L..49B} that star 110 Her is depleted in both Li and Be. We focus on two recent efforts: \citet{tucci2015} and \citet{boesgaard2022}. The first analyzed a small sample of solar-twins, and assumed a solar value of A(Be)=$1.38$ dex \citep[as per][]{asplund2009}. They found the Sun to be depleted in Be compared to their sample of 8 stars and found no signal of beryllium depletion as a function of age. However, the beryllium abundances they found seem unusually large. The second, and more recent effort, is a study analyzing beryllium abundances in a sample of approximately 50 one-solar-mass stars. They concluded that one of the most prominent explanations in the literature to explain the lithium abundance depletion, rotationally-induced extra mixing \citep[e.g.,][]{Deliyannis:1997ApJ...488..836D,Pinsonneault:2014ApJS..215...19P}, is not consistent with their data, as Be is not sufficiently depleted. \citet{boesgaard2022} claim that mixing due to gravity waves could be a potential explanation, as it does not deplete Be as much as Li. They also find a correlation between beryllium and metallicity, a possible indication of galactic nucleosynthesis. 

In this paper we build upon previous works and provide additional observational constraints to this unknown extra mixing mechanism observed in solar analogs by simultaneously analyzing lithium and beryllium abundances of a very restricted sample of solar twins. We select our targets based on the solar twin definition of \citet{Ramirez:2014A&A...572A..48R} in order to provide very strict constraints on the evolution and structure of stars that are like our most reliable testbed: the Sun. 

Our paper is organized as follows: In Section \ref{sec:data} we discuss our target selection and observations, data reduction is discussed in Section \ref{sec:data_reduction}, and we describe our abundance measurements and uncertainties in Section \ref{sec:data_analysis}. We discuss our findings in Section \ref{sec:discussion} and provide our concluding remarks in Section \ref{sec:conclusions}.

\section{Data} \label{sec:data}
\subsection{Target Selection and Observations}\label{sec:target_select}

First, we selected a sample of solar-twins by applying the photometric cuts proposed by \cite{galarza2021a} to the Gaia database. We then searched archival high-resolution, high-signal-to-noise (S/N) spectroscopic data. Using the cross-match between photometrically-selected solar-twins and spectra availability, we created  of bona-fide solar-twins with high precision photospheric stellar parameters. All of our targets have effective temperatures, surface gravities, metallicities, microturbulent velocities, as well as fundamental stellar parameters, namely masses and stellar ages. A description of our methodologies and the full sample is in preparation (Lorenzo-Oliveira et al. in prep). 

Our sample of solar twins only includes stars with $0.95 \le$ M(\sm) $\le 1.05$, $5677 \le$ \teff (K) $\le 5877$, $4.34 \le$ \logg (dex) $\le 4.54$, and $-0.1 \le$ \feh (dex) $\le +0.1$. We measured lithium abundances for all our sample and selected 23 stars with ages ranging from $0.5 \leq \tau_{\mathrm{iso}} \ \textrm{(Gyr)} \leq 8.0$. Our targets are all  constituted of nearby stars. The closest of them to the Sun (HIP 93858) being at a distance of $\sim17$ pc, and the farthest from the Sun (HIP 114615) at $\sim87$ pc. Distances were taken from \cite{Bailer-Jones:2021AJ....161..147B}.


All our targets were observed with the Ultraviolet and Visual Echelle Spectrograph \citep[UVES,][]{dekker2000}, at the VLT telescope under program ID 111.24VX.001. We used the blue configuration with the CD1 cross-disperser, centered at $346$ nm, and the 0.7" slit, yielding a resolving power of R$\approx55,000$, with an effective wavelength range from $302.5$ to $388.4$ nm. A log of our observations showing the full target list, along with the spectral signal-to-noise at $346$ nm is shown in Table \ref{tab:obs_log}. We also include in our analysis a solar spectrum, observed via the reflected light of the Juno asteroid (kindly provided by Jorge Meléndez, via private communication), using the same configuration in the blue, with the same resolving power and high S/N.

\begin{deluxetable*}{lccccc} 
\tabletypesize{\scriptsize} 
\tablecaption{Observation Log\label{tab:obs_log}} 
\tablewidth{0pt} 
\tablehead{
\colhead{Designation} & 
\colhead{R.A.} & 
\colhead{Decl.} & 
\colhead{UT Date} & 
\colhead{Exposure} & 
\colhead{S/N} \\ 
\colhead{} & 
\colhead{(deg)} & 
\colhead{(deg)} & 
\colhead{} & 
\colhead{Time (s)} & 
\colhead{$3460\sim\rm{\AA}$} 
}
\startdata
HIP 91287 & 279.305 & -25.674 & 2023-06-19 & 1530 & 249 \\
HIP 54582 & 167.554 & -7.391 & 2023-06-01 & 600 & 271 \\
HIP 118115 & 359.393 & -9.649 & 2023-07-24 & 2144 & 302 \\
HIP 79715 & 244.027 & -52.815 & 2023-05-09 & 1540 & 197 \\
HIP 85042 & 260.714 & -2.389 & 2023-06-06 & 380 & 236 \\
HIP 54582 & 167.554 & -7.391 & 2023-06-01 & 1023 & 376 \\
HIP 88595 & 271.345 & -21.674 & 2023-05-12 & 3232 & 311 \\
HIP 96334 & 293.791 & -69.978 & 2023-05-04 & 2146 & 254 \\
HIP 93858 & 286.718 & -37.813 & 2023-06-18 & 724 & 342 \\
HIP 50534 & 154.818 & -11.379 & 2023-05-16 & 2388 & 274 \\
HIP 96160 & 293.291 & -54.534 & 2023-05-06 & 2004 & 177 \\
HIP 64713 & 198.947 & -29.506 & 2023-06-18 & 1800 & 158 \\
HIP 59315 & 182.525 & -49.181 & 2023-05-28 & 2727 & 217 \\
HIP 116937 & 355.555 & -53.423 & 2023-07-20 & 2316 & 266 \\
HIP 74389 & 228.044 & -30.888 & 2023-06-10 & 1945 & 314 \\
HIP 85042 & 260.714 & -2.389 & 2023-06-06 & 768 & 371 \\
HIP 59532 & 183.117 & -3.082 & 2023-07-18 & 1530 & 288 \\
HIP 109821 & 333.666 & -41.387 & 2023-07-18 & 750 & 312 \\
HIP 114615 & 348.291 & -30.450 & 2023-07-31 & 3600 & 172 \\
HIP 62039 & 190.746 & -4.050 & 2023-07-18 & 2048 & 308 \\
HIP 116906 & 355.463 & -5.986 & 2023-07-05 & 1793 & 304 \\
HIP 54287 & 166.583 & -44.374 & 2023-05-21 & 1334 & 286 \\
HIP 108468 & 329.601 & -12.665 & 2023-06-06 & 1273 & 383 \\
HIP 95962 & 292.719 & -6.515 & 2023-05-12 & 1370 & 320 \\
HIP 91287 & 279.305 & -25.674 & 2023-07-08 & 1530 & 228 \\
HIP 54102 & 166.058 & -57.766 & 2023-05-06 & 3938 & 289
\enddata 
\end{deluxetable*}

\subsection{Data Reduction}
\label{sec:data_reduction}
We retrieved the UVES processed data products from the ESO archive. These include bias subtraction, flat field correction, optimized order extraction, S/N calculation, and barycentric velocity calculation. We then used the spectrum analysis tool \textit{iSpec}\footnote{\url{https://www.blancocuaresma.com/s/iSpec}} \citep{blanco2014,blanco2019} to perform radial velocity corrections and spectra normalization. For radial velocity corrections we performed a cross-correlation analysis using the VALD\footnote{http://vald.astro.uu.se/} linelist, which is integrated into \textit{iSpec}, to find a suitable radial-velocity (RV) correction for our abundance analysis. We then reduced the wavelength range of our RV corrected spectra to be between $306$ and $318$ nm. The only observable beryllium lines are the $3130.42$ and $3131.06$ \AA. 

\begin{figure}
 \includegraphics[width=\columnwidth]{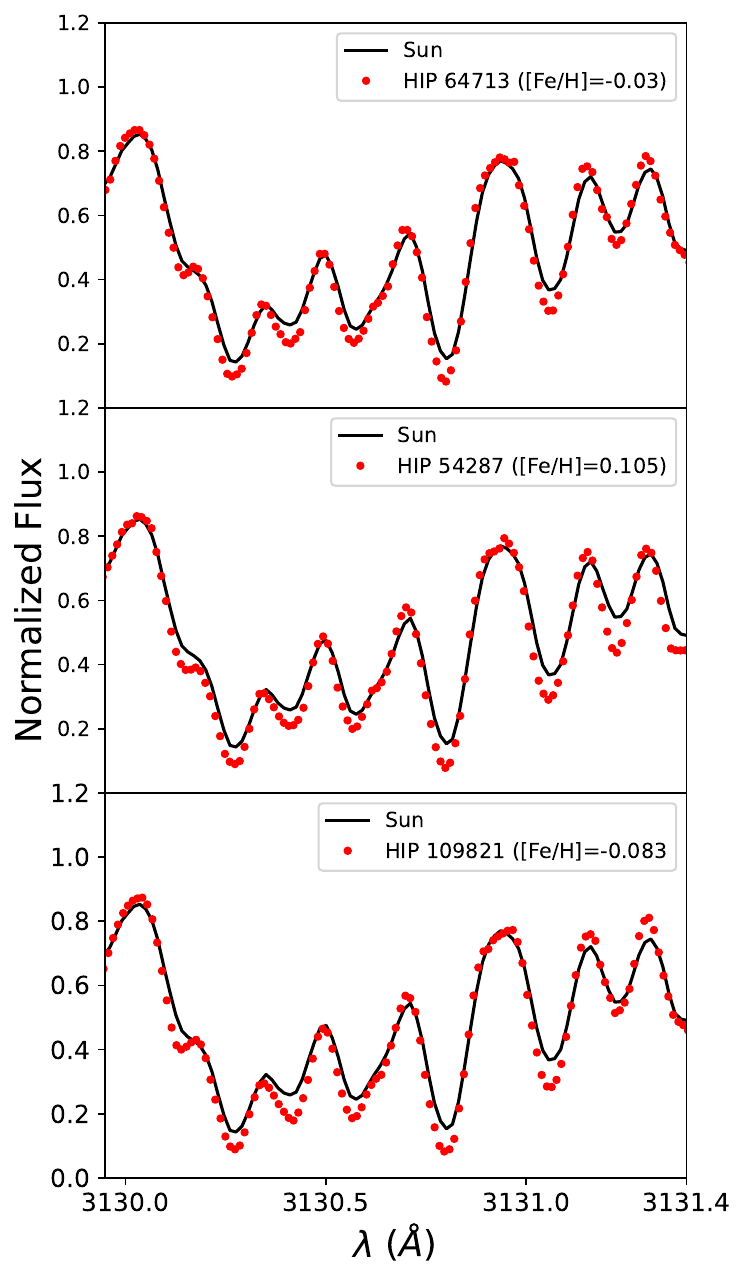}
 \centering
 \caption{Normalization example of the spectral region bracketing Be lines (3130 and 3131 \AA). In black we show the standard normalized solar spectrum, observed via the Juno asteroid, and in red three stars of our sample observed with the same instrument configuration. The same instrument configuration yielded the same resolution per wavelength, and the normalized spectrum is at the same level as that of the Sun.}
 \label{fig:normalization_comp}
\end{figure}


The spectral region containing the two beryllium atomic transition contains a multitude of different atomic and molecular transitions. Therefore, the normalization process can be prone to errors. To mitigate uncertainties in our measurements due to the normalization process, we normalize only our wavelength region containing the beryllium lines, and we do this using the previously normalized Juno spectrum in the same region as a template. In Figure \ref{fig:normalization_comp} we show the Juno normalized spectrum (black solid lines) overplot with the normalized spectrum of three stars with varying metallicities (HIP 64713, HIP 54287, and HIP 109821) to show that the continuum does not vary significantly in our sample. The process of normalizing all our science targets using an observed solar spectrum as a template guarantees that our differential analysis will contain the same (if any) systematics due to the normalization process. Therefore, we mitigate our relative uncertainties and guarantee that the relative abundances are precise.

To determine stellar parameters and lithium abundance, we employed stellar spectra collected by the HARPS \citep{Mayor:2003Msngr.114...20M} spectrograph. This instrument is mounted on the 3.6 m telescope at La Silla observatory (Atacama/Chile) and delivers a resolving power of $R = 115,000$ over a $\sim 380 - 690$ nm wavelength range. We retrieved reduced HARPS spectra from the ESO Archive\footnote{\url{http://archive.eso.org/wdb/wdb/adp/phase3_spectral/ form?phase3_collection=HARPS}}. For each one of our targets, we selected the twentieth highest SNR spectra under the following observational constraint: lowest possible air-mass ($\lesssim$ 2), and highest possible Moon angular separation ($\gtrsim$ 20 degrees) and SNR at 5500 \AA ($\gtrsim$ 100 pixels$^{-1}$). In those cases where eventually at least one of such conditions could not be met, we gradually relaxed each one of the observational constraints, ensuring a high quality master spectrum for each star. 

All spectra were normalized using IRAF's {\sc continuum} and doppler-shifted with \textit{dopcor} using the radial velocity inferred by the HARPS pipeline. In addition to the solar twins, we included a solar spectrum acquired with the HARPS instrument through observations of the asteroid Vesta to perform the differential analysis relative to the Sun.

\section{Data Analysis}
\label{sec:data_analysis}

\subsection{{Stellar Parameters}}
We determined atmospheric parameters through strictly differential analysis following \citet{Martioli:2023A&A...680A..84M} (see their section 3.4 and Appendix B) and Lorenzo-Oliveira et al. (in prep). Here, we provide a brief description of the process. Equivalent widths (EWs) of the atomic transitions of \ion{Fe}{1}, and \ion{Fe}{2}\citep{Melendez:2014ApJ...791...14M} were measured using an automatic Python script that fits Gaussian to a given line profile bracketed by a 6 \AA\ spectral window. The method is based on line-by-line equivalent width measurements between the Sun and the targets, choosing similar pseudo-continuum regions for both stars. 

To determine the stellar parameters (\teff, \logg, \feh, \vmic), we employed the automatic \qq\ (qoyllur-quipu) python code\footnote{\url{https://github.com/astroChasqui/q2}}. This code estimates iron abundances using the 2019 version of the local thermodynamic equilibrium (LTE) code MOOG \citep{sneden1973} with the Kurucz ODFNEW model atmospheres \citep{Castelli:2003IAUS..210P.A20C}. Then, \qq\ estimates the stellar parameters via spectroscopic equilibrium, which is a standard technique of iron line excitation and ionization equilibrium.

We estimated ages (here-forth $\tau_{\mathrm{iso}}$ for our inferred ages - based on isochrone fitting), masses, and radii of our sample through a Bayesian approach anchored on a different set of structural models, Yonsei-Yale evolutionary tracks \citep{Yi:2001ApJS..136..417Y, Demarque:2004ApJS..155..667D}. With the input parameters and its respective errors (\teff, \feh, [$\alpha$/Fe], \logg, \textit{Gaia} parallaxes, and $G$, $G_{\rm{BP}}$, and $G_{\rm{RP}}$ magnitudes), and posterior distribution functions derived from a proper likelihood marginalization along all possible evolutionary steps. Then, the likelihood is weighted by its respective mass (the \citealt{Salpeter:1955ApJ...121..161S} mass function) and metallicity \citep{Casagrande:2018IAUS..330..206C} priors to obtain the desired evolutionary parameter's median (50th percentile) and the $\pm1 \sigma$ intervals (16th to 84th percentile). For more details we refer to \cite{Diego:2019MNRAS.485L..68L}.

\subsection{Lithium Abundance calculations}
\label{sec:li_abnd}


To determine \li\ in our sample of solar twins, we followed the prescription given in \citet{Yana_Galarza:2016A&A...589A..17Y}, but for HARPS spectra. In summary, we determine macroturbulence velocity using the relation found by \citet{dosSantos:2016A&A...592A.156D} and projected rotational velocities by fitting the line profiles of five iron lines (6027.050, 6151.618, 6165.360, 6705.102 \AA,) and one nickel line (6767.772 \AA). Then, we synthesized the region around the asymmetric 6707.75 \AA\ Li line using the radiative transfer code \textsc{MOOG} \citep{sneden1973}, which assumes local thermodynamical equilibrium (LTE), with Kurucz model atmospheres \citep{Castelli:2003IAUS..210P.A20C}. We corrected all our non-LTE (NLTE) abundances using the grid from \citet{Lind:2009A&A...503..541L}.

\subsection{Beryllium Abundance calculations}
\label{sec:be_abnd}
As already pointed out, the near ultraviolet (just ultraviolet from now on) region containing the beryllium transitions is filled with different atomic and molecular transitions, which makes not only the normalization process difficult, but also complicates the process of generating the best match between the synthetic spectrum and the observed data. Further adding to this difficulty is the fact that not all the transitions in the region have atomic and molecular data measured in the laboratory, adding their uncertainties to the final systematic uncertainty of our data analysis. The beryllium transitions are also prone to departures from local thermodynamical equilibrium (non-LTE or NLTE departures) which can be significant \citep[e.g.,][]{korotin2022,Amarsi:2024A&A...690A.128A}.

As it is out of the scope of our study to produce a fully NLTE analysis of every spectrum, which would be time and computationally expensive, we decided to perform a differential approach to our analysis. We start by reproducing the observed solar spectrum with the current state-of-the-art solar abundances. We assume for all elements, except beryllium, the solar abundances from \citet{asplund2021}. For Be we performed two calibrations. First, we assumed the estimated abundance of \citet{korotin2022}. In our second iteration, we assumed the abundance from \citet{Amarsi:2024A&A...690A.128A}. \citet{korotin2022} used a 1D NLTE methodology estimated A(Be)$=1.32\pm0.05$ dex for the present-day solar beryllium abundance, fully compatible with the meteoritic values of A(Be)$=1.31\pm0.04$ dex. Based on 3D NLTE models, \citet{Amarsi:2024A&A...690A.128A} found slightly a lower abundance: A(Be)$=1.21\pm0.05$ dex.

We adopt a solar macroturbulent velocity of $\rm{V_{mac}}=3.5 \ \rm{km.s^{-1}}$, a rotational velocity of $\rm{vsin}i=1.9 \ \rm{km.s^{-1}}$ \citep{Bruning:1984ApJ...281..830B, Saar:1997MNRAS.284..803S}, a limb-darkening coefficient of $0.6$, and we calculate the instrumental broadening (assuming a Gaussian full-width half maximum - FWHM) at $3131$ \AA, for a resolving power of R$=55,000$, to be $0.06$.

Using the above parameters, we attempted to match our observed solar spectrum using the 2023 version of the 1D, LTE, radiative transfer code \textsc{MOOG}\footnote{https://moog-scat.readthedocs.io/en/latest/} \citep{sneden1973}. We highlight that the version of \textsc{MOOG} used in this study includes scattering, which is particularly important for shorter wavelengths. We adopt the MARCS 1D model atmospheres in our work, and our initial linelist was generated using the \textit{linemake}\footnote{https://github.com/vmplacco/linemake} code.

Our initial analysis did not provide a sufficiently good match between observed and synthesized spectrum, which we mostly attribute to two aspects: 1) The adopted solar abundances were mostly estimated using NLTE and, at times, 3D model atmospheres, opposed to our 1D, LTE analysis; 2) The shortcomings of analyzing such short wavelengths, namely normalization difficulties, and/or atomic/molecular line data of insufficient quality as well as the known issue of the missing oppacities in the Be region  \citep[e.g.,][]{Amarsi:2024A&A...690A.128A}.

We updated our linelist, by adjusting log$gf$ values and excitation potential values in order for our synthetic spectrum to match the observed solar spectrum with the known solar abundances. We updated the atomic data of the transitions for the lines available at \citet{korotin2022}, and then we adjusted the log$gf$'s and excitation potentials so that our synthesis would match the abundances of Be and the remaining elements in the synthesized region ($3129 - 3132$ \AA). By adjusting our linelist to ``solar log($gf$)s'' we created a linelist suitable for our analysis\footnote{Our full linelist is available by request.}. Our synthetic spectrum, matching the Sun, can be seen in Figure \ref{fig:solar_be}. We reiterate that we performed two instances of this same analysis. First, we updated our linelist to match the Be abundance reported by \citet{korotin2022} (A(Be)$=1.32$). Our second calibration matches the updated present-day solar abundance from \citet{Amarsi:2024A&A...690A.128A}. \citet{Amarsi:2024A&A...690A.128A} found a beryllium abundance that is somewhat lower: A(Be)$=1.21$, but they employed a 3D analysis, while \citet{korotin2022} did not. For details on the difference of the two methodologies, we refer the reader to the respective papers.

We proceed our analysis with the linelist that reproduces the present-day solar beryllium as calculated by \citet{Amarsi:2024A&A...690A.128A} (A(Be)$=1.21$). Then, we used our linelist to synthesize the beryllium region for the stars in our sample. Because we are first fine-tuning the linelist to match the solar properties, the resulting abundance zero-point shared by the solar twins is A(Be)$=1.21$. In other words, we adopt a differential approach in respect to the Sun to derive Be abundances for our sample stars. This approach takes advantage of the known solar spectroscopic similarity with solar twins to improve measurements of minute abundance differences compared to the Sun's absolute value. Regardless of the accuracy of the zero-point (A(Be)$=1.21$ or A(Be)$=1.32$), the precision on the relative abundances will be used throughout the analysis to study the depletion of Be as a function of time.

To synthesize the Be region in our sample of solar-twins we assumed a limb-darkening coefficient of $0.6$, and we calculate the instrumental broadening (assuming a Gaussian full-width half maximum - FWHM) at $3131$ \AA, for a resolving power of R$=55,000$, to be $0.06$. We calculated the macroturbulent velocity of each star following \citet{dosSantos:2016A&A...592A.156D}. Rotational velocity (\textit{vsin(i)}) was estimated along with the Be abundance at the region. We employed the 1D MARCS model atmospheres coupled to the LTE radiative transfer code MOOG, and the \textit{solar fine-tuned linelist} described above. In Table \ref{tab:stars} we present the Be abundances of our sample. 

\subsection{Uncertainty discussion}
The uncertainties in our stellar parameters are small. As a result of the precise analysis that is possible for solar-twins, the mean effective temperature uncertainty of our sample is $10$ K, the mean surface gravity uncertainty is $0.01$, and the mean metallicity uncertainty is $0.01$ dex. Propagating these uncertainties into our spectral synthesis analysis is difficult, as these small changes in the photospheric parameters cause virtually zero difference in the Be fitting. When attempting to quantify it, the uncertainty in beryllium abundance is between $0.01$ and $0.02$ dex, a value not discernible by our fitting process. The uncertainties due to our choice of macroturbulent velocity are similarly small. \cite{dosSantos:2016A&A...592A.156D} estimated a typical macroturbulent velocity uncertainties of $\sim0.1$ km.s$^{-1}$ for solar-twins. Therefore, we do not account for this source of error given its negligible impact on measured Be abundances, given the  resolution and SNR of our data.

\begin{figure}
 \includegraphics[width=\columnwidth]{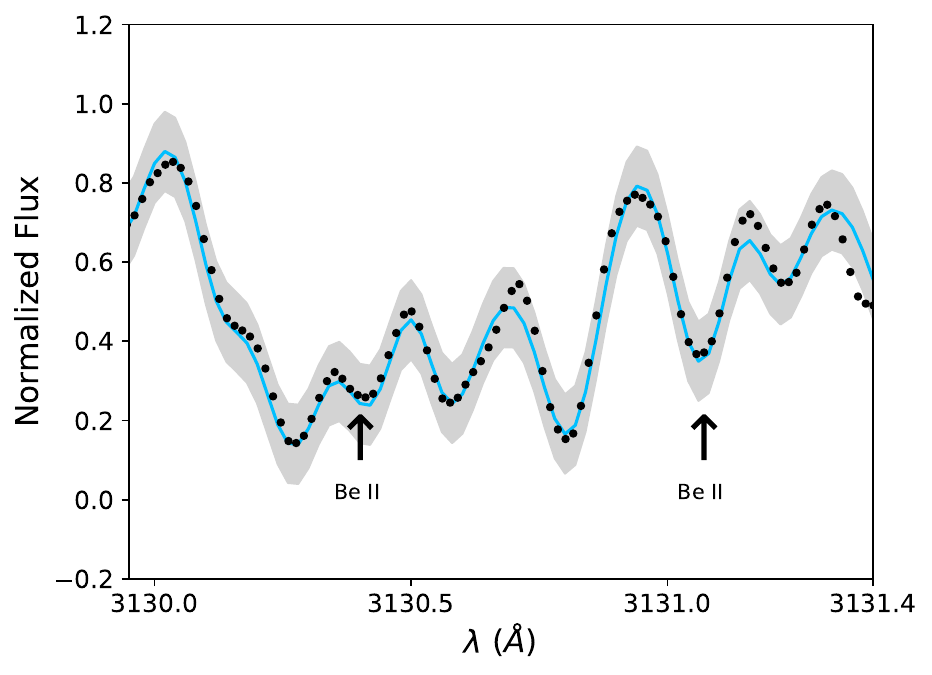}
 \centering
 \caption{The Solar spectrum in the beryllium region is shown as black dots, and our synthetic spectrum is shown in blue. The gray shaded area represents a $0.1$ dex uncertainty budget. We indicate the two \ion{Be}{2} transitions. As can be seen, our analysis yields uncertainties better than $0.1$ dex for the solar Be.}
 \label{fig:solar_be}
\end{figure}

On the other hand, there are several unknown uncertainties that affect our results. For example, the identification of several transitions in the UV region, and their associated atomic information. This is made clear by our methodology, in which we adjusted line information to match the observed solar spectrum. There are also normalization uncertainties, which can be significant, as no continuum region can be clearly identified. Therefore, a more reliable way to analyze the uncertainty budget in the beryllium abundance determination is directly via the spectral fitting itself. An example is seen in Figure \ref{fig:solar_be}, where the fitting of the solar beryllium is shown, along with a typical $\pm0.1$ uncertainty budget. By deriving uncertainties directly via the fitting process, we found the mean abundance uncertainty for the stars in our sample to be $\pm0.12$ dex in excellent agreement with the quoted uncertainty by \cite{boesgaard2022}. In terms of total error budget in Be abundance determinations, the authors claim that unknown sources of uncertainty may surpass the typical impact of stellar parameter errors.

\begin{deluxetable*}{lccccccc} 
\tabletypesize{\scriptsize} 
\tablecaption{Stellar Parameters and chemical abundances for our sample of stars. \label{tab:stars}} 
\tablewidth{0pt} 
\tablehead{
\colhead{Designation} & 
\colhead{\teff$^{a}$} & 
\colhead{\logg$^{a}$} & 
\colhead{\feh$^{a}$} & 
\colhead{Mass$^{a}$} & 
\colhead{$\tau_{\mathrm{iso}}^{a}$} & 
\colhead{A(Li)} & 
\colhead{A(Be)$^{b}$} \\ 
\colhead{} & 
\colhead{(K)} & 
\colhead{(cm s$^{-2}$)} & 
\colhead{(dex)} & 
\colhead{(M$_{\odot}$)} & 
\colhead{(Gyr)} & 
\colhead{(dex)} & 
\colhead{(dex)} \\ 
}
\startdata
HIP 50534 & $5729\pm10$ & $4.51\pm0.01$ & $+0.130\pm0.004$ & $1.05\pm0.02$ & $1.5\pm0.6$ & $1.61\pm0.04^{m}$ & $1.16\pm0.10$ \\ 
HIP 54102 & $5842\pm10$ & $4.50\pm0.01$ & $+0.018\pm0.005$ & $1.05\pm0.02$ & $1.5\pm0.7$ & $2.19\pm0.01{^c}$ & $1.17\pm0.10$ \\ 
HIP 54287  & $5724\pm10$ & $4.38\pm0.01$ & $+0.105\pm0.003$ & $1.02\pm0.03$ & $6.7\pm0.9$ & $1.91\pm0.01{^c}$ & $1.23\pm0.10$ \\ 
HIP 54582  & $5884\pm10$ & $4.28\pm0.02$ & $-0.065\pm0.005$ & $1.05\pm0.02$ & $6.8\pm0.4$ & $1.64\pm0.02{^c}$ & $1.16\pm0.10$ \\ 
HIP 59315  & $5647\pm13$ & $4.46\pm0.04$ & $+0.061\pm0.011$ & $0.98\pm0.02$ & $0.7\pm0.5$ & $2.85\pm0.02{^a}$ & $1.19\pm0.15$ \\ 
HIP 59532  & $5679\pm10$ & $4.37\pm0.01$ & $+0.139\pm0.004$ & $1.01\pm0.02$ & $7.1\pm1.0$ & $0.53\pm0.21{^m}$ & $1.17\pm0.15$ \\ 
HIP 62039  & $5749\pm10$ & $4.37\pm0.01$ & $+0.097\pm0.003$ & $1.03\pm0.02$ & $7.3\pm0.9$ & $0.81\pm0.19{^c}$ & $1.26\pm0.12$ \\ 
HIP 64713  & $5782\pm10$ & $4.38\pm0.01$ & $-0.030\pm0.003$ & $0.99\pm0.02$ & $5.1\pm0.7$ & $1.45\pm0.04{^c}$ & $1.14\pm0.10$ \\ 
HIP 74389  & $5842\pm10$ & $4.44\pm0.01$ & $+0.080\pm0.003$ & $1.06\pm0.02$ & $3.2\pm0.4$ & $2.09\pm0.01{^c}$ & $1.27\pm0.06$ \\ 
HIP 79715  & $5822\pm10$ & $4.38\pm0.01$ & $-0.024\pm0.003$ & $1.01\pm0.03$ & $6.7\pm0.7$ & $1.08\pm0.16{^c}$ & $1.25\pm0.13$ \\ 
HIP 85042  & $5699\pm10$ & $4.41\pm0.01$ & $+0.035\pm0.003$ & $0.99\pm0.02$ & $6.5\pm1.0$ & $0.53\pm0.13{^c}$ & $1.25\pm0.15$ \\ 
HIP 88595 & $5868\pm10$ & $4.43\pm0.01$ & $+0.074\pm0.010$ & $1.05\pm0.02$ & $4.9\pm0.5$ & $1.70\pm0.09{^a}$ & $1.35\pm0.10$\\ 
HIP 91287  & $5678\pm10$ & $4.43\pm0.01$ & $-0.031\pm0.003$ & $0.96\pm0.02$ & $5.6\pm1.0$ & $1.72\pm0.03{^a}$ & $1.20\pm0.15$ \\ 
HIP 93858  & $5654\pm10$ & $4.46\pm0.01$ & $+0.093\pm0.003$ & $1.00\pm0.02$ & $4.0\pm0.9$ & $0.25\pm0.10{^a}$ & $1.17\pm0.10$ \\ 
HIP 95962  & $5813\pm10$ & $4.41\pm0.01$ & $+0.029\pm0.002$ & $1.02\pm0.02$ & $5.7\pm0.7$ & $1.27\pm0.11{^c}$ & $1.26\pm0.10$ \\ 
HIP 96160  & $5797\pm10$ & $4.45\pm0.01$ & $-0.026\pm0.003$ & $1.01\pm0.02$ & $2.7\pm0.8$ & $1.75\pm0.03{^c}$ & $1.12\pm0.10$ \\ 
HIP 96334  & $5855\pm22$ & $4.51\pm0.04$ & $+0.094\pm0.014$ & $1.05\pm0.02$ & $0.6\pm0.5$ & $2.95\pm0.04{^a}$ & $1.27\pm0.15$ \\ 
HIP 108468  & $5842\pm10$ & $4.32\pm0.02$ & $-0.069\pm0.005$ & $1.01\pm0.02$ & $7.0\pm0.8$ & $1.13\pm0.14{^c}$ & $1.14\pm0.12$ \\ 
HIP 109821  & $5763\pm10$ & $4.31\pm0.01$ & $-0.083\pm0.003$ & $1.00\pm0.02$ & $8.0\pm0.6$ & $0.71\pm0.27{^c}$ & $1.15\pm0.10$ \\ 
HIP 114615  & $5824\pm10$ & $4.47\pm0.01$ & $-0.042\pm0.004$ & $1.02\pm0.03$ & $2.1\pm1.2$ & $1.89\pm0.02{^c}$ & $1.19\pm0.10$ \\ 
HIP 116906  & $5797\pm10$ & $4.38\pm0.01$ & $+0.000\pm0.002$ & $1.02\pm0.02$ & $6.7\pm0.9$ & $0.78\pm0.22{^c}$ & $1.22\pm0.10$ \\ 
HIP 116937 & $5658\pm10$ & $4.46\pm0.01$ & $+0.025\pm0.003$ & $0.98\pm0.03$ & $3.1\pm0.8$ & $0.73\pm0.07{^a}$ & $1.10\pm0.10$ \\ 
HIP 118115 & $5798\pm10$ & $4.26\pm0.01$ & $-0.028\pm0.004$ & $1.02\pm0.03$ & $7.8\pm0.6$ & $0.96\pm0.09{^c}$ & $1.25\pm0.15$
\enddata 
\tablecomments{$^{a}$ Lorenzo-Oliveira et al (in prep), $^{c}$ \citet{Carlos:2019MNRAS.tmp..667C}, $^{m}$ \citet{Martos:2023MNRAS.522.3217M}, $^{b}$ This work.}
\end{deluxetable*}



\begin{figure}
 \includegraphics[width=\columnwidth]{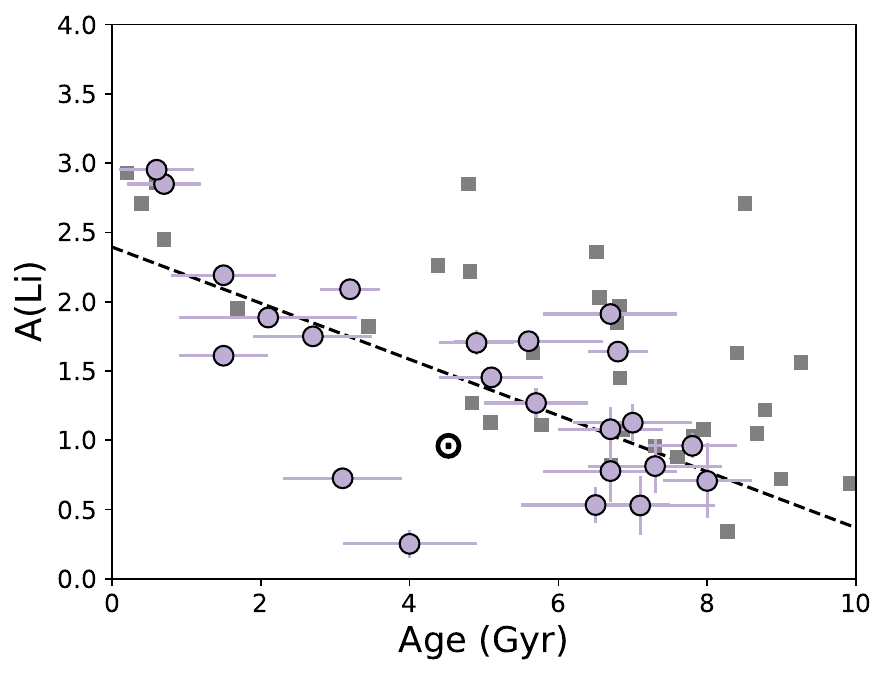}
 \centering
 \caption{Lithium abundances of our sample of solar-twins as a function of stellar age. Lithium measurements of our sample as represented as purple circles, the Sun is plotted as a black solar standard symbol, and the gray squares are the lithium abundances of stars from \cite{boesgaard2022} that satisfy our solar-twin criteria.}
 \label{fig:li_age}
\end{figure}

\section{Discussion}
\label{sec:discussion}

The lithium abundances of the stars in our sample have been previously derived in different studies, in particular by \cite{Carlos:2019MNRAS.tmp..667C} and \cite{Martos:2023MNRAS.522.3217M}. We re-derived all abundances in an effort to homogenize our analysis with the stellar parameters we found. In Figure \ref{fig:li_age} we show our lithium abundances as a function of age. Our abundances are displayed in purple and the gray squares represent the measurements from \citet{boesgaard2022} for the stars in their sample that satisfy our solar twin criteria. The dashed line indicates the linear fit to our data (excluding the \citet{boesgaard2022} stars). Unsurprisingly, as we are using the same linelist and methodology for deriving lithium as previous works, we find good agreement between our abundances and those from previous studies. The scatter we found in our lithium abundances is quite tipycal of other studies analyzing lithium in solar twins \citep[e.g., ][]{Carlos:2016A&A...587A.100C, Rathsam:2023MNRAS.525.4642R, Martos:2023MNRAS.522.3217M}. The linear correlations between lithium and stellar ages we find are 

\begin{equation}
    \mathrm{A(Li)}=-0.20\pm0.05\times \tau_{\mathrm{iso}} + 2.39\pm0.26
    \label{eq:lithium_age}
\end{equation}

\begin{equation}
    \tau_{\mathrm{iso}}=-3.02\pm0.46\times \mathrm{A(Li)} + 9.52\pm0.76
    \label{eq:age_lithium}
\end{equation}

with ages ($\tau_{\mathrm{iso}}$) in Gyr. Although our sample is smaller, and we do have two stars with low lithium abundances (HIP 93858, and HIP 116937), Equations \ref{eq:lithium_age} and \ref{eq:age_lithium} agree with previous works from \citet{Carlos:2019MNRAS.tmp..667C}, \citet{Martos:2023MNRAS.522.3217M}, and \citet{Rathsam:2023MNRAS.525.4642R}. The latter found $\mathrm{A(Li)=-0.27\pm0.02\times \mathrm{\tau_{iso}} + 2.65\pm0.11}$. If we remove the two low lithium outliers (HIP 93858, and HIP 116937), the relationship between lithium and age we find 

\begin{equation}
    \mathrm{A(Li)}=-0.23\pm0.03\times \tau_{\mathrm{iso}} + 2.64\pm0.19    
\end{equation}

agrees even better with the above mentioned studies. However, we found somewhat higher uncertainties in our fits, due to our reduced sample size. It is important to highlight that lithium in our selected sample has the standard behavior of main-sequence depletion in solar-twins.

We briefly mention our outliers. The star with the highest lithium abundance for its age (HIP 54287 - A(Li)$=1.91\pm0.01$, with an age of $\tau_{\mathrm{iso}} = 6.7\pm0.9$ Gyr), had already been previously identified as an outlier by \citet{Carlos:2019MNRAS.tmp..667C}. The second one, HIP 54582 (A(Li)$=1.64\pm0.02$, with an age of $\tau_{\mathrm{iso}} = 6.8\pm0.4$ Gyr), has a detected brown dwarf companion of 20.7 M$_{\mathrm{JUP} }$ \citep{Feng:2022ApJS..262...21F}. The $\sim$3.5 Gyr-old stars showing very low lithium content are candidates for a further investigation in the near future since they might resemble the recently reported young solar-twin with the lowest lithium abundance detected so far, HIP 8522 \citep{Yana:2024arXiv241017590Y}.

\begin{figure}
 \includegraphics[width=\columnwidth]{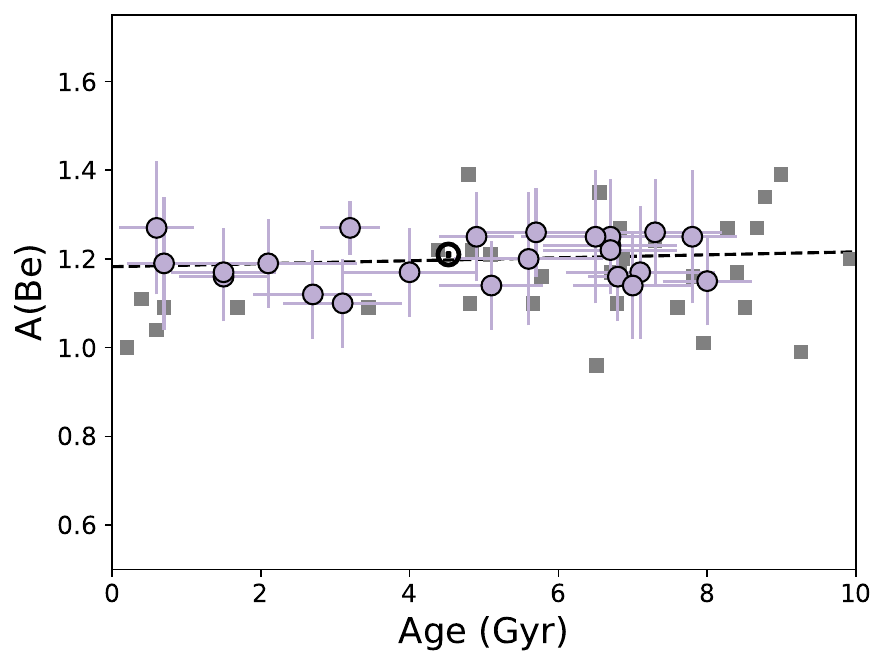}
 \centering
 \caption{Beryllium abundances of our sample of solar-twins as a function of stellar age. The Sun is plotted as a black solar standard symbol, and the gray squares are the beryllium abundances of stars from \cite{boesgaard2022} that satisfy our solar-twin criteria. Our data do is consistent with a flat beryllium distribution as a function of stellar age, consistent with no beryllium depletion.}
 \label{fig:be_age}
\end{figure}

In Figure \ref{fig:be_age}, we show the abundances of beryllium as a function of stellar age. We also include, as gray squares, a sample of stars from \cite{boesgaard2022} that satisfy our solar-twin criteria. \cite{boesgaard2022} found a solar abundance of A(Be) $=1.23$ dex in their analysis of the Sun (using a sky spectrum), very similar to the abundance adopted in this work. There is good agreement between their analysis and the 1D LTE analysis performed by \cite{takeda2011} as well. 

There is a total of 34 stars that meet our criteria in \cite{boesgaard2022}, and there is an abundance dispersion of $0.11$ dex if we exclude the outliers. In our sample, composed of 23 stars, the abundance dispersion is $0.05$ dex. We attribute the difference in scatter to the higher S/N of our data and to our calibration based on the most recent available solar abundances. It is important to say, though, that the differences are small and within measurement uncertainties. Our scatter is comparable to the $0.04$ dex scatter found by \citet{tucci2015}, which also did a differential study. However, their sample is considerably smaller (8 stars). \citet{tucci2015} found beryllium abundances that are considerably higher (up to A(Be)$=1.52$ for one of their target stars). We do not include any of their measurements, as there seems to be too large disagreements, which we cannot explain, compared to our data and to other high precision studies of beryllium \citep[e.g.,][]{takeda2011,boesgaard2022}.

We fit a linear regression model (dashed black line) to our beryllium abundances as a function of age and find 

\begin{equation}
    \mathrm{A(Be)}=0.003\pm0.004\times \tau_{\mathrm{iso}} + 1.18\pm0.02
\end{equation}

\begin{equation}
    \tau_{\mathrm{iso}}=7.13\pm9.95\times \mathrm{A(Be)} - 3.71\pm11.94,
\end{equation}

for $\tau_{\mathrm{iso}}$ in Gyr. Although the slope found in \citet{boesgaard2022} data is slightly higher than ours ($0.01$ vs 
$0.003$), the correlation between Be abundance and age is flat, and there is not any discernible beryllium depletion in solar-twins through their main-sequence lifetime. Our results cannot probe pre-main-sequence Be burning, which would require a different observational approach, but we show that the mixing mechanism responsible for dredging lithium and burning it during the main-sequence is unable to dredge material deep enough in the star to reach temperatures capable of Be burning ($3.5\times 10^6$ K).

Current standard models can't reproduce beryllium and lithium burning. To check their behavior we did a very simple experiment with the open-source 1D stellar evolution code \textsc{MESA} \citep[Modules for Experiments in Stellar Astrophysics][]{paxton2011, paxton2013, paxton2015, paxton2018, paxton2019, jermyn2023}. Following \citet{Sevilla:2022MNRAS.516.3354S}, we modified the initial metallicity (Z) in the MESA inlists they used to predict lithium evolution in solar-type stars. Our simple model was calibrated to match present-day solar properties (bolometric luminosity, radius, and temperature) by adopting initial Z=0.016 to account for diffusion effects. Also, to reproduce the secular lithium evolution trend in solar-type stars the model, in its non-rotating version, includes thermohaline mixing \citep{Brown:2013ApJ...768...34B}, and elemental diffusion \citep{paxton2018} only after ZAMS. For more details on the model, we refer to \citet{Sevilla:2022MNRAS.516.3354S}. For beryllium, we used the same toy inlists, but we assumed the protosolar Be abundance of A(Be) $= 1.40$ \citep{Lodders:2021SSRv..217...44L}. While the data shows that the initial beryllium abundance at the zero-age-main-sequence (ZAMS) should be the same as the current abundance (A(Be)$\sim1.21$), our toy model did not include ZAMS burning. Therefore, it only reaches the current value at approximately 10 Gyrs. In our toy model, in the 10 Gyr range probed by our data, beryllium decreases from A(Be)$=1.40$ to A(Be)$=1.23$. Even if we account for our data scatter and uncertainties, the data does not show a $0.17$ dex decrease in abundance over the same period. Therefore, a standard solar evolution model depletes far more beryllium during the main-sequence than what is observed in our sample of solar-twins. What our simple exercise shows is that, indeed, current models cannot simultaneously explain what is observed in lithium and beryllium abundances in solar-like stars. 

\begin{figure}
 \includegraphics[width=\columnwidth]{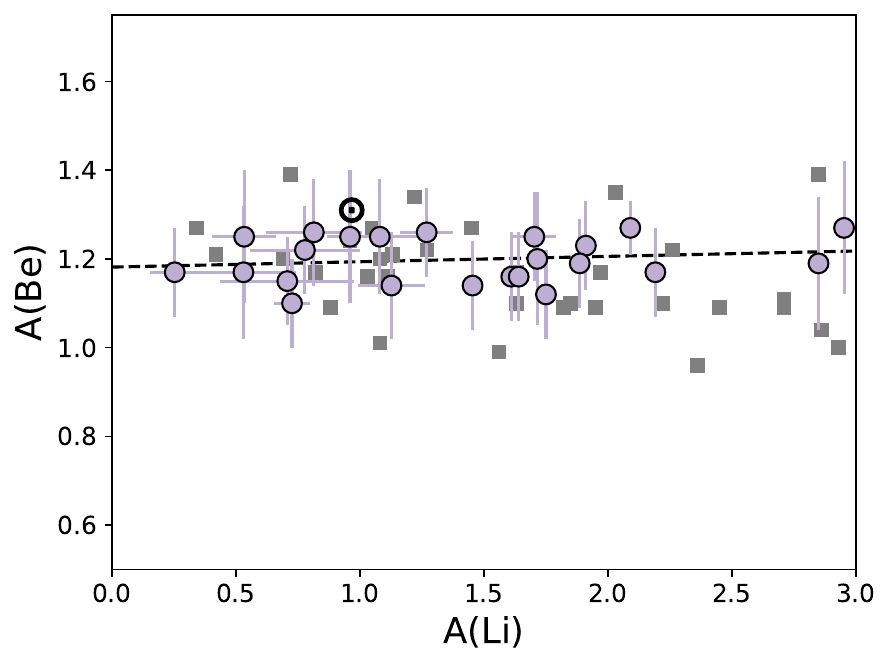}
 \centering
 \caption{Beryllium abundances as a function of lithium abundances. Symbols are the same as in Figure \ref{fig:be_age}. We do not find any significant slope in the data, which excludes rotationally-induced extra mixing models as the main driver of the observed lithium depletion as a function of stellar age.}
 \label{fig:li_be}
\end{figure}

As previously pointed-out, \citet{boesgaard2022} discussed how their data compares to predictions from models favoring rotational mixing as the mechanism to deplete lithium and beryllium, but the data does not support it. Rotational models deplete too much, up to $0.4$ dex, beryllium during the main sequence \citep{Deliyannis:1990ApJ...365L..67D}, and the data does not support such level of depletion. Furthermore, these rotational models \citep[e.g.][]{Deliyannis:1997ApJ...488..836D} show a slope between A(Be) and A(Li) of about 0.4 for F stars, with a decrease for cooler stars, such as those in our sample. In Figure \ref{fig:li_be} we show the relation between beryllium and lithium in our sample. We found a negligible slope of $0.01$. In their study \citet{boesgaard2022} also found a negligible slope of $-0.008$. Neither slopes are significant and they agree within uncertainties. This is further evidence that rotationally-induced mixing models burn significantly more Be than observed. \citet{Dumont:2021A&A...646A..48D} and \citet{Dumont:2021A&A...654A..46D} rotationally-induced mixing models for F and G stars at the main-sequence do not find as much Be burning as previous models, but still predict more Be burning than observed in our work (see their Figure 4, and our Figure \ref{fig:li_be}).

We find that our results, for $1 \ \mathrm{M_{\odot}}$ stars, are better represented by the models of \citet{Xiong:2007ChA&A..31..244X}. Their models reproduce the typical lithium burning while no beryllium burning is observed for their $1 \ \mathrm{M_{\odot}}$ models during the main sequence (see their Figure 1). The convective settling models by \citet{Andrassy:2015A&A...579A.122A} also presents itself as a good alternative, as it also does not burn as much beryllium at the temperature range discussed in this paper. As discussed in \citet{boesgaard2022}, mixing models that include gravity waves can also be a good explanation, as they do not burn Be as much as Li.

\begin{figure}
 \includegraphics[width=\columnwidth]{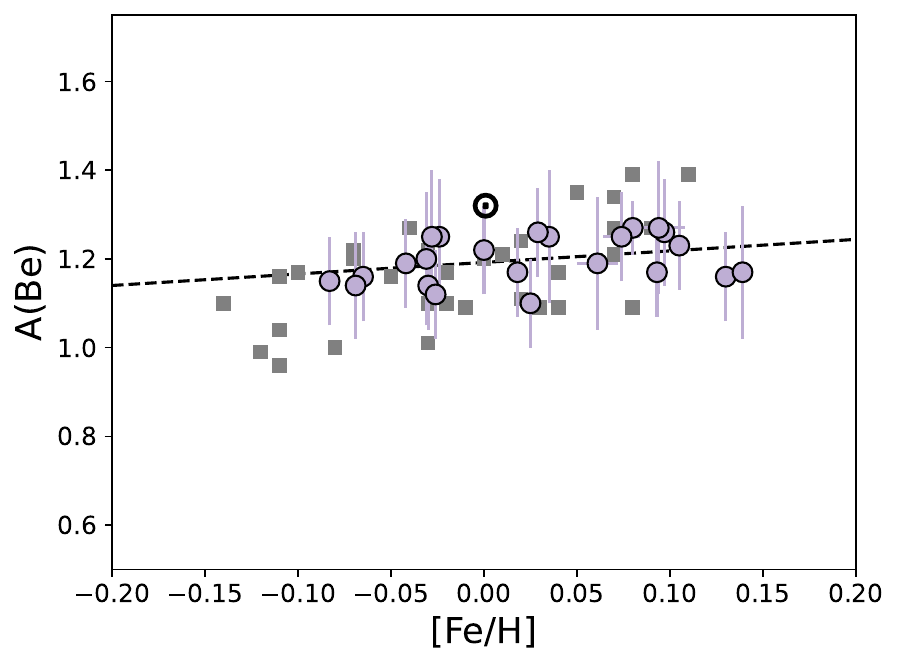}
 \centering
 \caption{Beryllium abundances of our sample of solar-twins as a function of stellar metallicity. The symbols are the same as Figure \ref{fig:be_age}. The observed positive slope is indicative of galactic beryllium nucleosynthesis via cosmic ray spallation processes.}
 \label{fig:be_feh}
\end{figure}

Interestingly, we do find a clear trend with metallicity even in our very narrow metallicity range ($-0.15 \le \rm{[Fe/H]} \le +0.15$), where 

\begin{equation}
    \mathrm{A(Be)}=0.26\pm0.16\times \rm{[Fe/H] + 1.19\pm0.01}, 
\end{equation}

a very discernible slope of $0.26$. We show this result in Figure \ref{fig:be_feh}. This slope is comparable to the slope of $0.34$ found by \citet{boesgaard2022}, although their sample extended to metallicities as low as [Fe/H]$=-0.49$ dex.

The only stable beryllium isotope that is observable in stellar photospheres is $^9$Be. This isotope cannot be produced in stellar interiors, and its nucleosynthesis is limited to spallation reactions (i.e., breaking apart of heavier nuclei by high-energy cosmic rays). Therefore, as there is no observable depletion of beryllium as a function of age, our results are an excellent tracer of spallation nucleosynthesis for models of the galactic chemical evolution, corroborating the conclusions by \citet{boesgaard2022}.

We found no trend with effective temperature or stellar mass in our sample. This is not surprising, as the range in both are very small, and any trend is unlikely to be observable.

\citet{Amarsi:2024A&A...690A.128A} present-day solar photospheric beryllium abundance measurement (A(Be)$=1.21\pm0.05$) is $0.10$ dex lower than the best CI chondrite measurement of A(Be)$=1.31\pm0.04$ dex \citep{Lodders:2021SSRv..217...44L}. \citet{Amarsi:2024A&A...690A.128A} points out that if one assumes the initial solar beryllium abundance to be the one from the meteorites, A(Be)$=1.31$, beryllium has been depleted by $22\pm11\%$ on top of any effects of atomic diffusion, in tension with Standard Solar Models. Non-Standard Solar Models, on the other hand, predict excessive burning, as we pointed by our work as well. Based on our toy model described above, at the Solar age, the Sun would have depleted $\approx0.05$ dex of Be, roughly in agreement with the scenario presented in Figure 9 of \citet{Amarsi:2024A&A...690A.128A}. The Non-Standard model they presented, which includes effects of angular momentum transport, predicts the abundance of A(Be)$=1.21$ at an age of $\sim1.5$ Gyr. The solar-twins in our sample with estimated ages of $\sim1.0$ Gyr already have beryllium abundances on the level of A(Be)$=1.21$, which means that any Be burning must have happened at an earlier stage, and quickly, without considerable burning as the star evolves. This Be burning is therefore unrelated to the well-known lithium depletion as a function of time, observed in main sequence solar-twins.

As was also concluded by \citet{Amarsi:2024A&A...690A.128A}, our work suggests that we still need further modelling developments to study not only Li burning, but Be as well. On the observational side, further work is needed at younger stars, closer to the ZAMS, to constraint at which point Be abundances are depleted to the mean observed value of A(Be)$\sim1.21$ in the Sun and solar-twins. It is also important to identify the unknown blend at the Be line \citep{Amarsi:2024A&A...690A.128A} and corroborate current best estimate of the solar Be abundance.

\section{Conclusion}
\label{sec:conclusions}

In this paper we describe the analysis of lithium and beryllium abundances in a sample of solar-twins with a comprehensive age range. We design a new and improved linelist (available upon request) for 1D LTE analysis of beryllium abundances that reproduce the most recent 3D non-LTE Be abundance measurements. 

We find no variation of beryllium abundance as a function of stellar age, indicating that the extra mixing mechanism responsible for the observed lithium burning cannot bring material deep enough to burn beryllium as well. We compare our observations to a toy model and show that standard assumptions, scaled to reproduce solar lithium, burn beryllium in excess when compared to our observations. Compared to non-standard models, our observations seem to favor convective ovsershooting and convective settling over rotationally-induced mixing models.

Our results indicate that the beryllium depletion observed in the sun (from A(Be)$=1.31$ from the meteorites to A(Be)$=1.21$ from the present-day solar photosphere) likely happens before the first Gyr. More work is still needed, at younger stars, to help constrain how this initial burning takes place and why it stops, both from observational and modeling perspectives.

As previously found, we see an increase in the beryllium abundances as a function of metallicity, possibly indicative of galactic beryllium production via cosmic ray spallation and in novaes.

\section*{Acknowledgments}
Henrique Reggiani acknowledges the support from NOIRLab, which is managed by the Association of Universities for Research in Astronomy (AURA) under a cooperative agreement with the National Science Foundation. Jhon Yana Galarza acknowledges support from a Carnegie Fellowship. Diego Lorenzo-Oliveira acknowledges the support from CNPq (PCI 301612/2024-2). Sofia Covarrubias acknowledges support from the Carnegie Astrophysics Summer Student Internship Program (CASSI) and the Ralph M. Parsons Foundation. J.C. acknowledges support from the Agencia Nacional de Investigación y
Desarrollo (ANID) via Proyecto Fondecyt Regular 1231345, and from ANID
BASAL project CATA2-FB210003. We also thank Joshua D. Simon and Andrew McWilliam for the useful discussions about this project.

\vspace{5mm}
\facilities{VLT:Kueyen, ESO:3.6m, \textit{Gaia}}

\software{
\textsc{numpy} \citep{van_der_Walt:2011CSE....13b..22V}, 
\textsc{matplotlib} \citep{Hunter:4160265}, 
\textsc{pandas} \citep{mckinney-proc-scipy-2010}, 
\textsc{iraf} \citep{Tody:1986SPIE..627..733T}, 
\textsc{iSpec} \citep{Blanco:2014A&A...569A.111B, Blanco:2019MNRAS.486.2075B}, 
\textsc{Kapteyn} Package \citep{KapteynPackage}, 
\textsc{moog} \citep{Sneden:1973PhDT.......180S}, 
q$^{\textsc{2}}$ \citep{Ramirez:2014A&A...572A..48R}.
}

\bibliography{be}{}
\bibliographystyle{aasjournal}

\end{document}